\documentclass[journal]{IEEEtran}
\usepackage{cite}
\usepackage{enumerate}
\usepackage{amsmath}
\usepackage{amsfonts}
\usepackage{mathrsfs}
\usepackage{amssymb}
\usepackage{cases}
\usepackage{stmaryrd}
\usepackage{mathrsfs}
\usepackage{bbm}
\usepackage[dvips]{graphicx}
\usepackage{algorithm}
\usepackage{algorithmic} 
\usepackage{multirow}
\usepackage{graphicx}
\usepackage{bm}
\usepackage{color}
\usepackage{setspace} 
\usepackage[acronym]{glossaries}

\usepackage[colorlinks]{hyperref}
\usepackage{threeparttable}
\usepackage{multirow}
\usepackage{setspace}

\ifCLASSINFOpdf
\else
\fi

\newacronym{PLS}{PLS}{physical layer security}
\newacronym{IRS}{RIS}{Reconfigurable Intelligent Surfaces}
\newacronym{MIMO}{MIMO}{Multiple-Input Multiple-Output}
\newacronym{SER}{SER}{symbol error rate}
\newacronym{DFRC}{DFRC}{dual-functional radar-communications}
\newacronym{LFM}{LFM}{orthogonal linear frequency modulation }
\newacronym{NLOS}{NLOS}{non-line-of-sight }
\newacronym{AOA}{AOA}{angle-of-arrival }
\newacronym{AOD}{AOD}{angle-of-departure }
\newacronym{LOS}{LOS}{line-of-sight}
\newacronym{PAR}{PAPR}{peak-to-average-power ratio}
\newacronym{AO}{AO}{alternating optimization}
\newacronym{Eve}{Eve}{eavesdropper}
\newacronym{BPSK}{BPSK}{binary phase shift keying}
\newacronym{CI}{CI}{constructive interference}
\newacronym{QoS}{QoS}{quality-of-service}
\newacronym{bs}{BS}{base station}
\newacronym{STARS}{STAR-RIS}{simultaneously transmitting and reflecting reconfigurable intelligent surfaces}
\newacronym{dfrc}{DFRC}{dual functional radar-communications }
\newacronym{SINR}{SINR}{signal-to-interference-plus-noise ratio}
\newacronym{admm}{ADMM}{alternating direction method of multipliers}
\newacronym{MM}{MM}{majorization-minimization }
\newacronym{sdr}{SDR}{semidefinite relaxation}
\newacronym{BS}{BS}{base station}
\newacronym{UE}{UE}{user equipment}
\newacronym{MRT}{MRT}{maximum ratio transmission}
\newacronym{ZF}{ZF}{zero-forcing}
\newacronym{SNR}{SNR}{signal-noise-ratio}
\newacronym{FAS}{FAS}{Fluid Antenna System}
\newacronym{SFAMA}{s-FAMA}{slow fluid antenna multiple access}
\newacronym{FFAMA}{f-FAMA}{fast fluid antenna multiple access}
\newacronym{SIR}{SIR}{signal-to-interference ratio}
\newacronym{CSI}{CSI}{channel state information}
\newacronym{NMSE}{NMSE}{normalized minimum square error}
\newacronym{MAE}{MAE}{masked autoencoders}
\newacronym{TSP}{TSP}{ travelling salesman problem}
\newacronym{LSTM}{LSTM}{Long Short-Term Memory}
\newacronym{DOF}{DoF}{degrees of freedom}
\newacronym{A2C}{A2C}{advantage actor and critic}
\newacronym{DNN}{DNN}{deep neural network}
\newacronym{ISAC}{ISAC}{integrated sensing and communication}
\newacronym{DRL}{DRL}{deep reinforcement learning}
\newacronym{AI}{AI}{artificial intelligence}
\newacronym{NOBP}{NOBP}{number of observed ports}
\newacronym{llm}{LLM}{Large Language Model}
\newacronym{mllm}{MLLM}{Multimodal Large Language Model}
\newacronym{RAG}{RAG}{retrieval-augmented generation}
\newacronym{LLM-MA}{LLM-MA}{Large Language Model-based multi-agent}
\newacronym{CKM}{CKM}{Channel Knowledge Map}
\newacronym{HPO}{HPO}{hyperparameter optimization }
\newacronym{LHH}{LHH}{Language Hyper-Heuristics}
\newacronym{ReEvo}{ReEvo}{reflective evolution}
\newacronym{GA}{GA}{genetic algorithm}
\newacronym{CoT}{CoT}{Chain of Thought}
\newacronym{MARL}{MARL}{multi-agent reinforcement learning}
\newacronym{eSpark}{eSpark}{Evolutionary action Space Reduction with Knowledge}
\newacronym{DoF}{DoF}{degrees-of-freedom}
\newacronym{AP}{AP}{access point}
\newacronym{fpga}{FPGA}{Field-Programmable Gate Array}
\newacronym{asic}{ASIC}{Application-Specific Integrated Circuit}
\newacronym{gpu}{GPU}{Graphics Processing Unit}
\newacronym{tpu}{TPU}{Tensor Processing Unit}
\newacronym{NOMA}{NOMA}{Non-Orthogonal Multiple Access}

\newtheorem{Remark}{Remark}
\usepackage{array}
\makeatletter
\newcommand{\thickhline}{%
    \noalign {\ifnum 0=`}\fi \hrule height 1pt
    \futurelet \reserved@a \@xhline
}
\newcolumntype{"}{@{\hskip\tabcolsep\vrule width 1pt\hskip\tabcolsep}}
\makeatother

\definecolor{chao}{rgb}{0,0,1}
\begin{document}

%\title{Leveraging Large Language Models for the Design and Optimization of Fluid Antenna Systems
\title{Large Language Model Empowered Design of\\Fluid Antenna Systems: Challenges, Frameworks, and Case Studies for 6G
%	\thanks{
%		The work of C. Wang is supported in part by the National Natural Science Foundation of China under Grants 62371357 and 61801518, in part by the International Postdoctoral Exchange Fellowship Program under Grant PC2021060,
%		in part by the China Postdoctoral Science Foundation under Grant 2020M683428;
%		the work of K. K. Wong is supported  by the Engineering and Physical Sciences Research Council (EPSRC) under Grant EP/W026813/1;
%		the work of Z. Li is supported by the National Science Fund for Distinguished Young Scholars of China under Grant 61825104.}
	}
%\title{A Complete Survey for Fluid Antenna Systems and the Road Ahead for Wireless Communications}

\author{Chao Wang,
            Kai-Kit Wong,
            Zan Li,
            Liang Jin, and 
            Chan-Byoung Chae

\vspace{-.36in}

\thanks{C. Wang and Z. Li are  with the Integrated Service Networks Lab, Xidian University, Xi'an 710071, China.}
\thanks{K. K. Wong is with the Department of Electronic and Electrical Engineering, University College London, Torrington Place, U.K. He is also affiliated with the Yonsei Frontier Laboratory, Yonsei University, Seoul, 03722, South Korea (e-mail: kai-kit.wong@ucl.ac.uk).}
\thanks{L. Jin is with Information Engineering University, Zhengzhou, 450001, China (e-mail: liangjin@263.net).}
\thanks{C.-B. Chae is with the School of Integrated Technology, Yonsei University, Seoul, 03722 South Korea (e-mail: cbchae@yonsei.ac.kr).}

}

\markboth{Submitted to IEEE Wireless Communications, 2025}%
{6G Wireless System}

\maketitle

\begin{abstract}
The Fluid Antenna System (FAS), which enables flexible Multiple-Input Multiple-Output (MIMO) communications, introduces new spatial degrees of freedom for next-generation wireless networks. Unlike traditional MIMO, FAS involves joint port selection and precoder design, a combinatorial NP-hard optimization problem. Moreover, fully leveraging FAS requires acquiring Channel State Information (CSI) across its ports, a challenge exacerbated by the system's near-continuous reconfigurability. These factors make traditional system design methods impractical for FAS due to nonconvexity and prohibitive computational complexity. While deep learning (DL)-based approaches have been proposed for MIMO optimization, their limited generalization and fitting capabilities render them suboptimal for FAS. In contrast, Large Language Models (LLMs) extend DL's capabilities by offering general-purpose adaptability, reasoning, and few-shot learning, thereby overcoming the limitations of task-specific, data-intensive models. This article presents a vision for LLM-driven FAS design, proposing a novel flexible communication framework. To demonstrate the potential, we examine LLM-enhanced FAS in multiuser scenarios, showcasing how LLMs can revolutionize FAS optimization.
\end{abstract}


\IEEEpeerreviewmaketitle

\vspace{-2mm}
\section{Introduction}
Multi-antenna systems have undergone transformative advancements to meet escalating wireless demands. 
%The introduction of \gls{MIMO} systems marked a significant breakthrough, utilizing spatial multiplexing and diversity to enhance  link reliability. 
The breakthrough adoption of massive \gls{MIMO} marked a pivotal milestone, serving as a cornerstone of 5G networks through its revolutionary improvements in spectral efficiency and energy performance.
The advent of \gls{FAS} now represents the next paradigm shift, introducing fully reconfigurable antenna architectures with dynamic tuning capabilities \cite{Tutorial_FAS}. FAS achieves unmatched channel adaptation by enabling quasi-continuous antenna reconfiguration, surpassing traditional discrete antenna systems~\cite{VirtualFAS}. When integrated with \gls{AI}-driven optimization, this technology is expected to form the foundation for 6G networks, characterized by ultra-reconfigurability and exceptional capacity.

Depending on the implementation, FAS, sometimes dubbed ``movable antenna'', is a novel reconfigurable antenna array capable of switching between multiple positions (``ports") within a defined area. Recent studies have increasingly explored FAS designs for various scenarios.
%Recently, \gls{FAS} has been integrated with \gls{MIMO} to enhance the multi-stream transmission performance by designing spatial channels through optimizing the ports \cite{Port_selection}. 
%There are an increasing number of works studied the \gls{FAS} design for different scenarios. 
For example, 
 %work in \cite{Port_selection} proposed an online learning framework to optimize port selection for slow fluid antenna multiple access.    
\gls{FAS}-liberated \gls{ISAC} systems  have been explored in \cite{ISAC_FAS}. While deep learning (DL) techniques are increasingly being adopted for FAS design \cite{AI_FAS}, they suffer from heavy reliance on large, costly-to-acquire training datasets and limited generalization across diverse wireless channel conditions, key challenges hindering practical deployment.

{\color{chao}Large Language Models (LLMs) are undergoing transformative advancements, demonstrating unprecedented capabilities across multiple domains, particularly in optimization problem generation and solving, creating a closed-loop paradigm for automated decision-making \cite{Niyato1, Niyato2}. In problem generation, LLMs synthesize realistic optimization scenarios by learning historical data patterns, autonomously formulating objective functions and constraints. For problem-solving, LLMs excel in complex, high-dimensional optimization landscapes where traditional methods struggle. Specifically, LLM-enhanced heuristic design leverages advanced reasoning, knowledge retrieval, and adaptive learning to efficiently tackle intricate problems. By interpreting unstructured problem descriptions, LLMs decompose tasks into manageable sub-problems, propose tailored heuristics (e.g., genetic algorithms), and dynamically refine parameters (e.g., mutation rates, exploration-exploitation trade-offs) using real-time feedback. These capabilities make LLMs uniquely suited for next-generation wireless communication design, particularly in FAS \cite{FAS-LLM1}. 

Unlike conventional DL models, LLMs leverage multimodal knowledge processing to incorporate domain-specific insights from technical literature, improving its generalization performance. Their strong generalization enables adaptation to diverse wireless environments, while natural language processing facilitates interpretable, physics-aware optimization, addressing DL's ``black-box'' limitations. Furthermore, LLMs enhance FAS design through three key mechanisms \cite{FAS-LLM1}: 
\begin{enumerate}
\item Research synthesis (aggregating technical knowledge); 
\item Optimization suggestions (proposing design improvements), and 
\item Automated parameter tuning (fine-tuning system configurations).
\end{enumerate}
 A novel LLM-empowered port selection technique \cite{Port_LLM} exemplifies this by exploiting LLMs' pattern recognition and reasoning abilities to optimize complex sequences, advancing human-AI collaborative engineering.}

\begin{figure*}[!th]
	\centering
	\includegraphics[width=5.3in]{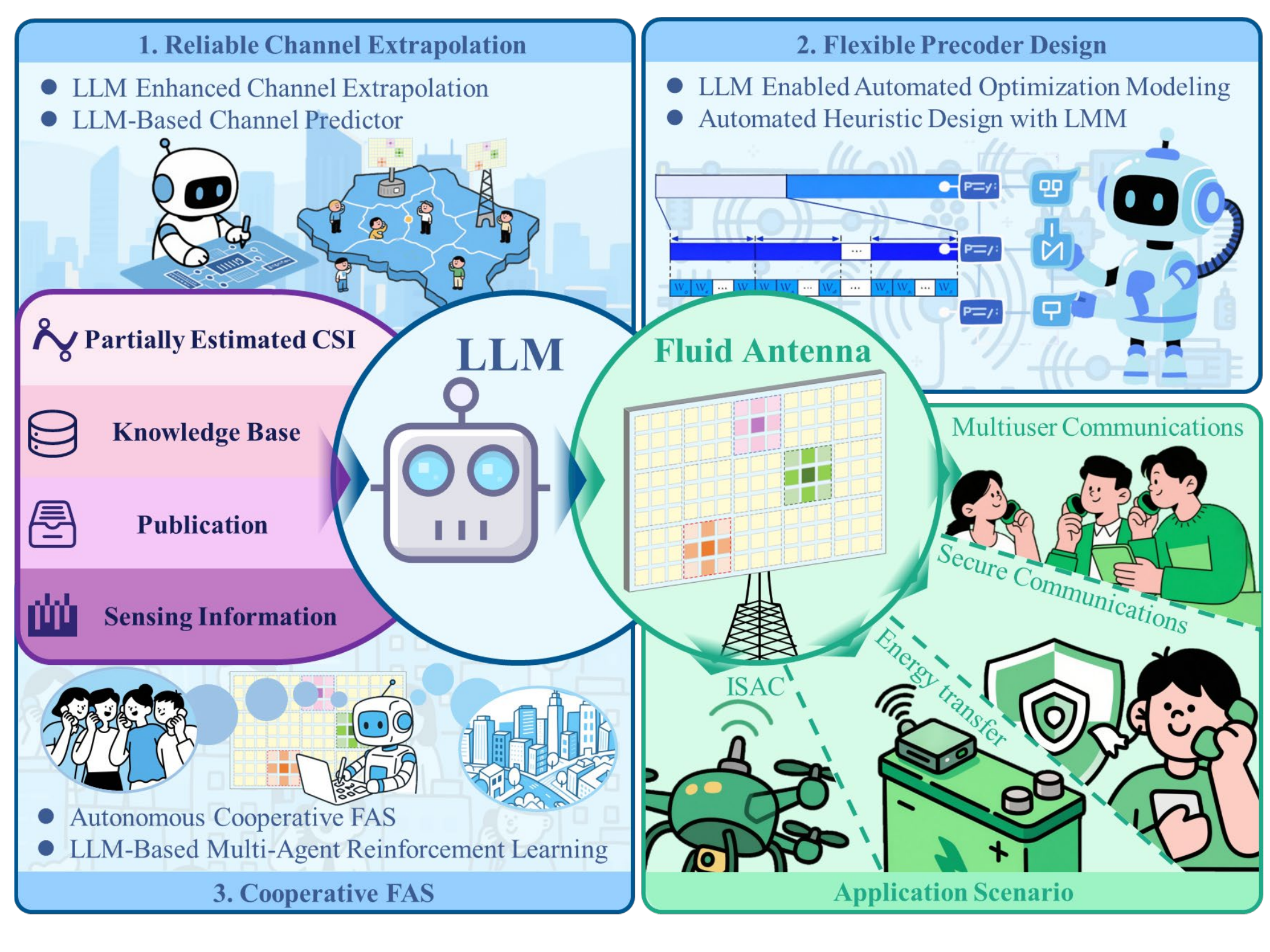}
	\caption{The interplay of \gls{FAS} and \gls{llm}.}\label{Interplay}
\end{figure*}

This work presents an \gls{llm}-powered intelligent \gls{FAS} design framework. Section II outlines the mission-critical objectives and fundamental challenges of intelligent FAS. In Section III, we propose an LLM-driven FAS design methodology, detailing: (1) key LLM architectural principles and (2) their novel applications in channel extrapolation, optimal port selection, and cooperative FAS optimization. Section IV validates the framework through a multiuser communication case study, demonstrating LLM-enabled automation and significant performance gains. Finally, Section V concludes with key insights and future research directions.

\vspace{-2mm}
\section{An Overview of Intelligent FAS}\label{sec:prelim}
\gls{FAS}, enabling position- and shape-flexible antenna arrays, have emerged as a promising technology for future 6G networks. However, their advantages are accompanied by significant design challenges. Below, we outline the key challenges, particularly in: Channel estimation, flexible precoder design, and cooperative \gls{FAS} optimization.

\subsection{Reliable Channel Extrapolation}
%In \gls{FAS}, \gls{CSI} is essential for port selection and precoder design. However, estimating the \gls{CSI} for all possible port combinations within limited timeframes presents a significant challenge. This makes reliable channel extrapolation a critical requirement for effective \gls{FAS} design.
1) \textbf{Mission:}
Unlike traditional multi-antenna systems, \gls{FAS} utilizes a significantly larger number of ports, especially when antennas are capable of moving within a predefined space. This reconfigurability introduces complexities that make precise channel estimation extremely challenging, if not impossible. Consequently, implementing \gls{FAS} requires a redesigned transmission protocol that replaces traditional \gls{CSI} estimation with channel extrapolation \cite{VirtualFAS}. This approach leverages the estimated \gls{CSI} of a small subset of ports to predict the \gls{CSI} for all available ports. %Since channels across different ports may be non-reciprocal, spatial channel reciprocity can be exploited for extrapolation due to the shared scattering environment.
Traditional channel extrapolation approaches typically employ linear or nonlinear techniques, which estimate values outside a known range by extending observed trends. However, the effectiveness of these methods depends on the accuracy of the extrapolated mathematical model, which often fails to exploit the spatial correlation patterns among multiple ports. While DL is increasingly used to predict CSI by leveraging spatial correlations, it lacks generalization ability and often requires retraining for new tasks. Therefore, FAS requires a robust and efficient channel extrapolation module with strong scalability to accurately extrapolate \gls{CSI} based on the estimated ones.
%Additionally, \gls{CKM} has recently emerged as a key technology for providing location-specific statistical channel knowledge, enabling environment-aware communications. Given the large number of invisible ports in \gls{FAS}, there is a need to design a low-complexity channel extrapolation approach by effectively exploiting \gls{CKM}.

2) \textbf{Challenges:}
To enable accurate channel prediction, the extrapolation module of \gls{FAS} must be redesigned, as it involves a significantly larger number of invisible ports compared to traditional antenna arrays, presenting new challenges \cite{VirtualFAS}.

(a). \textit{Robust Extrapolation Approaches:} Designing robust channel extrapolation methods for FAS is challenging due to the large number of inactive ports and the complex dependencies in the multi-port CSI \cite{Tutorial_FAS}. 
%Essentially, the channel extrapolation problem is a time-series forecasting task, specifically one involving long sequence forecasting. The goal is to predict the \gls{CSI} of invisible ports over an extended period based on observed data. 
First, capturing long-term dependencies is difficult because the influence of observable port data tends to decay or become corrupted over large spatial separations, and conventional DL models often fail to preserve critical information across extended sequences.
Second, the \gls{CSI} across different ports frequently exhibits intricate, non-linear correlations that are poorly modeled by traditional neural network architectures.

(b). \textit{Efficient \gls{CKM} Exploitation:} {\color{chao}The \gls{CKM} provides prior information for channel extrapolation in \gls{FAS}, which can reduce system overhead and improve prediction accuracy \cite{CKM_ChannelPrediction}. However, integrating \gls{CKM} as external knowledge to enhance channel extrapolation presents significant challenges.
First, due to the time-varying nature of wireless channels, \gls{CKM} may become outdated, limiting its effectiveness for \gls{CSI} prediction. 
%This raises the first challenge: how to extract useful and relevant information from \gls{CKM}.
Second, DL as a data-driven approach, requires retraining networks to account for the varying correlation patterns of \gls{CSI} across different ports to ensure prediction accuracy. However, designing customized deep neural networks for each correlation pattern is impractical due to the high design complexity. }
%This leads to the second challenge: designing an adaptive deep neural network capable of adjusting to different statistical \gls{CSI} or \gls{CKM}.

\subsection{Flexible Precoder Design}
%The reconfigurable antenna ports introduce new spatial \gls{DoF} to \gls{FAS}, enabling flexible precoder design. However, this necessitates the joint optimization of port selection and precoder designa a nonconvex problem that further complicates the development of a universal model for diverse applications.

1) \textbf{Mission:}
Compared to traditional \gls{MIMO}, \gls{FAS} offers greater spatial \gls{DoF} that enables more flexible waveform design. Specifically, port selection establishes beneficial spatial channels while precoder optimization maximizes system performance by fully utilizing these channels \cite{Tutorial_FAS}. However, the joint design constitutes an NP-hard problem due to the combinatorial nature of port selection, resulting in prohibitively high computational complexity for \gls{FAS} implementations.
This challenge necessitates novel, efficient precoder design approaches to effectively exploit the spatial \gls{DoF} in \gls{FAS}. Moreover, \gls{FAS} must accommodate diverse applications having varying requirements and mathematical models. Different applications demand distinct channel configurations based on their specific service requirements. For instance, 
%In \gls{ISAC}, antenna positions should align sensing and communication channels to enable dual-function signal reuse.
in multiuser communications, positions should optimize channel orthogonality to minimize interference.
Therefore, to fully realize the potential of \gls{FAS}, the system requires automated modeling capabilities that can adapt to different scenarios and application requirements.

2) \textbf{Challenges:} Designing flexible precoders for \gls{FAS} requires an intelligent optimization framework to automate port selection and precoder design across diverse wireless applications, presenting significant implementation challenges \cite{AI_FAS}. 

(a). \textit{Automatic Problem Modeling:} Integrating \gls{FAS} into mobile communications demands the construction of distinct problem models tailored to diverse applications. To realize this, an intelligent FAS should feature automatic problem-modeling capabilities, enabling it to extract actionable insights from real-world operational data and synergistically integrate them with domain expertise to formulate  design problem models. However, critical challenges persist: 1) Sensing-to-model translation: Seamlessly converting sensory inputs into precise problem models remains nontrivial. {\color{chao}Although \cite{Niyato2} proposed leveraging \gls{RAG} to automate problem modeling, their work was limited in scope, considering only restricted scenarios, access protocols, and optimization goals. More critically, they did not account for multi-modal information in the automated formulation of optimization problems.}
2) Knowledge integration: Effectively leveraging existing expert knowledge (e.g., from IEEE Xplore and Google Scholar) to refine models, particularly for channel characterization and joint port selection-precoder optimization, poses additional complexity.
 
%Meanwhile, ongoing research on fluid antennas has yielded extensive theoretical results, which can inform the optimization models for FAS implementation.
%Intelligent FAS systems should leverage existing expert knowledge (e.g., from IEEE Xplore and Google Scholar) to refine system modelsa such as channel modeling and joint port selection-precoder optimization, which presents another significant challenge.

(b). \textit{Nonconvex Joint Optimization:} Joint port selection and precoder optimization in FAS pose a nonconvex and computationally intractable problem \cite{ISAC_FAS}. While DL offers a data-driven solution method, its poor generalization capability often leads to overfitting on limited training data and failure on unseen configurations. Consequently, developing an efficient optimization algorithm for this joint design remains a significant challenge.
Furthermore, port selection constitutes a combinatorial optimization problem, where the feasible region grows exponentially with the number of available ports. 
%This exponential scaling renders traditional learning-aided optimization methods ineffective, as deep learning lacks problem-specific heuristics to prune irrelevant search regions. 
%Instead, it performs blind exploration, incurring high computational costs without guarantees of optimality. 
Thus, achieving a scalable joint design algorithm for FAS presents another critical research challenge.

\subsection{Cooperative Multi-Agent \gls{FAS}}
%Cooperative multi-agent \gls{FAS} has extended single-agent \gls{FAS}, through using multiple \gls{FAS}s to accomplish  the task collaboratively. However,  the increasing number of \gls{FAS}s brings the surge in the number of spatial \gls{DoF} making the cooperative system design challenging. 
1) \textbf{Mission.}
Future networks will shift from traditional cellular systems to cell-free massive MIMO, where distributed \gls{AP}s cooperatively serve users. When \gls{AP}s are equipped with fluid antenna, cooperative \gls{FAS} design becomes essential, yet its vast feasible region renders centralized optimization impractical due to prohibitive computational complexity \cite{Tutorial_FAS}. %Specifically, the antenna ports of multiple \gls{FAS} devices must be jointly optimized to construct effective cooperative channels. However, this necessitates extensive information exchange among FAS units, leading to a combinatorial optimization problem with an intractably large solution space.
Besides, when multiple APs collaborate on a task, the task can be decomposed into sub-tasks distributed across them. However, centralized allocation remains challenging due to the configurable channel vectors of FAS.
To address these challenges, a multi-agent FAS framework should autonomously decompose tasks based on mission requirements, enabling distributed optimization where each agent independently adjusts its FAS configuration using local sensing data. 
%Yet, traditional distributed optimization methods struggle with the non-convex joint design problem inherent in FAS. Instead, a multi-agent learning approach is needed, allowing FAS devices to learn cooperative strategies directly from environmental observations.

2) \textbf{Challenges:} Developing efficient cooperative multi-agent \gls{FAS}s requires an autonomous multi-agent learning framework to overcome the prohibitive computational complexity of joint optimization, presenting fundamental design challenges.

(a) \textit{Autonomous Multi-Agent Cooperation}: Designing cooperative multi-agent systems for \gls{FAS} demands the automation of three critical functions: wireless environment perception, task decomposition, and adaptive strategy design \cite{AI_FAS}.
First, environmental perception poses a key challenge: how to fuse distributed sensor data from multiple \gls{FAS}s into a unified situational model that informs cooperative decision-making and multi-agent learning.
Second, task decomposition must dynamically adapt to environmental constraints (e.g., node locations, spatial channels). The core difficulty lies in translating perceptual knowledge into modular subproblems that enable efficient multi-agent coordination.
Third, achieving intelligent cooperation requires human-like cognitive abilities, such as expert-guided reasoning and meta-learning from past interactions, to continuously refine collaboration strategies.

(b) \textit{Cooperative Multi-Agent Learning}. %The large action spaces in \gls{FAS} design complicate the development of cooperative multi-agent learning algorithms, as the joint action space grows exponentially with the number of agents. 
The development of cooperative multi-agent learning algorithms for collaborative \gls{FAS} faces three fundamental challenges \cite{MARL_Training}:
1) Curse of Dimensionality: The prohibitively large joint action space of multiple \gls{FAS}s makes exploration and optimization computationally intractable.
2) Exploration Difficulties: The likelihood of discovering beneficial joint actions decreases significantly.
3) Coordination Complexity: Identifying optimal action combinations across multiple \gls{FAS}s becomes harder, exacerbated by the non-stationarity of the wireless environment.
%Policy Representation: High-dimensional action spaces demand complex function approximators, which are often unstable.
To address these challenges, advanced techniques, such as efficient exploration strategies are essential for enabling effective cooperation among \gls{FAS}s.

%\vspace{-2mm}
\section{\gls{llm} liberating FAS for 6G}\label{sec:design}
The powerful capabilities of \gls{llm}, including cross-modal generation, information retrieval, logical reasoning, and task planning,
can help alleviate the design complexity caused by uncertain antenna positions in \gls{FAS} and enhance its practical application. 
%In particular, the cross-modal generation capability allows \gls{FAS} to leverage fused data from multimodal sensors, enabling a proactive design approach that addresses reliability and latency challenges caused by incoming blockages and other dynamic factors in wireless scenarios.
%Besides, the information retrieval and logical reasoning capabilities can enable automated optimization modeling and problem solving for different applications of \gls{FAS}. 
%In addition, the task planning capability can enable intelligent cooperative \gls{FAS} through dynamic task replanning and resource allocation.

\subsection{Overview of \gls{llm}}
\subsubsection{Multimodal \gls{llm}}
\gls{mllm}, which refers to \gls{llm}-based models capable of receiving, reasoning, and generating multimodal information, has garnered increasing attention since the release of GPT-4 \cite{mllm}. A typical \gls{mllm} comprises three core modules: a modality encoder, a pre-trained \gls{llm}, and an interface module that bridges the two.
Specifically, the encoder acts as the system's ``eyes and ears'', compressing multimodal inputs (e.g., images, audio) into a compact representation. 
%However, since \gls{llm} can only process text, it is essential to introduce an interface that converts multimodal information into a format that can be understood by \gls{llm}.
%Generally, we can implement the learnable interface using token-level and feature-level fusion, or alternatively, use an expert model to achieve the interface without requiring training.
An \gls{llm}, acting as the 'brain' to process compressed multimodal information, can be a pre-trained model embedded with extensive world knowledge, possessing strong understanding and reasoning capabilities.
\subsubsection{AI-Based Automated Problem Modeling and Solving}
%The formulation and solution of optimization problems in specialized fields require domain-specific expertise, mathematical proficiency, and technical coding skills, creating significant challenges for both novice users and experienced professionals. However, recent advancements in \gls{llm} technology offer promising potential to reshape how engineering problems are modeled and solved. 
%\gls{llm}, in particular, excel at natural language understanding and generating logically structured responses, allowing users to express problems in intuitive, non-technical terms. 
%While \gls{llm} are trained on extensive, general-purpose datasets that may lack in-depth expertise in specific engineering domains, \gls{RAG} can enhance their capabilities \cite{RAGLLM}. By integrating \gls{llm}' language-processing strengths with specialized knowledgea such as a database of research papers or industry-specific datasetsa RAG can significantly improve their performance in specialized fields.
%In conclusion, \gls{llm} is expected to replace human experts in problem modeling and algorithm development, thereby facilitating the design of wireless systems.
Optimizing problems in specialized fields requires domain expertise, mathematical proficiency, and coding skills, presenting challenges for both novices and professionals. Recent advances in \gls{llm} could revolutionize how engineering problems are modeled and solved. {\color{chao}Although LLMs are trained on broad datasets with limited domain-specific depth, \gls{RAG} can improve their accuracy by incorporating specialized knowledge (e.g., research papers or industry datasets). Specifically, RAG enables LLMs to leverage expert knowledge for automated problem modeling without human intervention. Furthermore, LLMs can intelligently select or customize algorithms, automating algorithm development and streamlining wireless system design.}
\subsubsection{\gls{llm} Multi-Agent Collaborative Task-Solving}
%Intelligent agents are computational entities endowed with autonomy, reactivity, and communication capabilities.
%\gls{llm} has emerged as a powerful framework for constructing these agents, comprising four core components: 1) Knowledge Base: Integration with external knowledge bases allows agents to leverage domain-specific information for context-aware responses. 2) Tools: Interfaces enabling agents to process tasks (e.g., data retrieval, calculations, or system operations).
%3) Memory: A module for storing, retrieving, and updating past interactions and learned behaviors. 4) Model: The \gls{llm} itself, which interprets natural language inputs and converts them into executable commands.
%Compared with a single \gls{llm}-empowered agent, \gls{LLM-MA} systems leverage the specialized skills of multiple agents, have demonstrated significant potential in tackling complex problems \cite{llm_multiagent}. 
%\gls{LLM-MA} systems break complex tasks into smaller, manageable components, with multiple agents assigned different roles that collaborate to tackle the challenge. These agents learn from each other and make decisions based on a combination of shared knowledge and expertise. 
%Additionally, \gls{LLM-MA} systems leverage the powerful planning and decision-making capabilities of \gls{llm} to enable efficient distributed cooperation by facilitating communication, sharing knowledge, coordinating tasks, and adapting to new information. 

Intelligent agents are autonomous, reactive computational entities with communication capabilities \cite{Niyato2}. \gls{llm} provides a framework for building such agents, structured around four components:
1) Knowledge Base: Integrates domain-specific data for context-aware responses;
2) Tools: Interfaces for task execution (e.g., data retrieval, calculations);
3) Memory: Stores and retrieves past interactions for adaptive learning;
4) Model: The LLM core, translating natural language into executable commands.
In LLM-MA (Multi-Agent) systems, agents decompose complex tasks into subtasks, assume specialized roles, and collaborate by sharing knowledge and expertise.

\subsection{Intelligent \gls{FAS} Empowered by \gls{llm}}
We introduce the intelligent \gls{FAS} empowered by \gls{llm}, concentrating on designing its physical layer techniques\footnote{\color{chao}
Since FAS performance depends heavily on the propagation environment, constructing a  CSI database is critical for training LLMs. Our approach begins by leveraging publicly available datasets like  \href{https://www.deepmimo.net/}{DeepMIMO} to establish baseline MIMO channel models, followed by environment-specific CSI generation using ray-tracing simulations (e.g., Wireless InSite) or MATLAB's 5G Toolbox. For real-world validation, we can collect over-the-air measurements via software-defined radios (SDRs) such as USRP or LimeSDR while dynamically adjusting fluid antenna positions. The raw CSI data is then preprocessed into structured embeddings, converting complex channel matrices into time-series sequences and labeling them with optimal antenna configurations, before being used to train the LLM for FAS optimization.}, as shown in Fig. \ref{Interplay}. 
\subsubsection{Reliable Channel Extrapolation}
The challenges in channel extrapolation for \gls{FAS} arise from the dynamic nature of spatial correlations and the high number of unobserved ports. To address these challenges, \gls{llm} and \gls{CKM} can be leveraged to construct the prior information of spatial correlations. This prior information can be exploited to optimize hyper-parameters, thereby improving channel extrapolation performance. 
%Additionally, traditional channel prediction algorithms face significant limitations in accuracy, practicality, and scalability.  This is primarily because they rely on fixed-step \gls{CSI} sequences for \gls{CSI} prediction. It is well known that \gls{llm} has demonstrated exceptional pattern recognition and reasoning capabilities when handling complex sequences. These strengths can be harnessed to predict the next "token" in \gls{CSI}, offering a more robust solution \cite{llm_CSIprediction}.

\textbf{\gls{llm} Enhanced Channel Extrapolation.} \gls{CKM} is a site-specific database that provides location-specific channel information to facilitate  \gls{CSI} acquisition \cite{CKM_ChannelPrediction}. However, due to positioning errors and time-varying scatterers, \gls{CKM} alone cannot fully capture instantaneous \gls{CSI}. Despite this limitation, \gls{CKM} serves as valuable prior information, reducing system overhead and enhancing estimation performance. Notably, there are both common and distinct components between the \gls{CSI} derived from \gls{CKM} and the true \gls{CSI}. 
%This is because two nearby locations may share some scatterers while also having unique ones. 
\gls{CKM} can provide \gls{CSI} for paths with matching angles and delays, allowing us to focus on estimating \gls{CSI} for paths that differ in these aspects. {\color{chao}By leveraging \gls{CKM} as external knowledge, the integration of \gls{llm} and \gls{RAG} enables \gls{FAS} to extract useful \gls{CSI} from \gls{CKM}, thereby improving \gls{CSI} estimation \cite{RAGLLM}. Additionally, DL offers a data-driven, flexible, and efficient approach to channel estimation, making it a promising solution for the complex and dynamic environments of modern wireless networks. However, the limited generalization of DL models across different scenarios remains a significant challenge, hindering the practicality of learning-based CSI estimation. To address this, \gls{llm} enable automated \gls{HPO}, which adjusts hyperparameters governing the model's structure and learning algorithms. This adaptability allows learning-based estimation approaches to be tailored to diverse scenarios, enhancing their effectiveness \cite{zhang2023using}.}

\textbf{\gls{llm}-Based Channel Predictor.} Channel prediction is fundamentally a time-series forecasting problem. Traditional deep neural networks for time-series analysis often employ compact architectures with a limited number of learnable parameters, which constrains their ability to address complex scenarios. In contrast, \gls{llm}s  possess powerful modeling and generalization capabilities that could be applied to time-series forecasting. However, \gls{llm}s are primarily designed for natural language processing (NLP), creating a significant challenge in aligning wireless communication data with NLP data formats. To bridge this gap, the disparity between \gls{CSI} (Channel State Information) and natural language data can be addressed by developing specialized preprocessor, embedding, and output modules \cite{LLM_channelestimation}. These modules enable the transformation of \gls{CSI} prediction into a time-series problem, leveraging \gls{llm}s' inherent strength in processing sequential data.

\begin{figure*}[!t]
	\centering
	\includegraphics[width=6.3in]{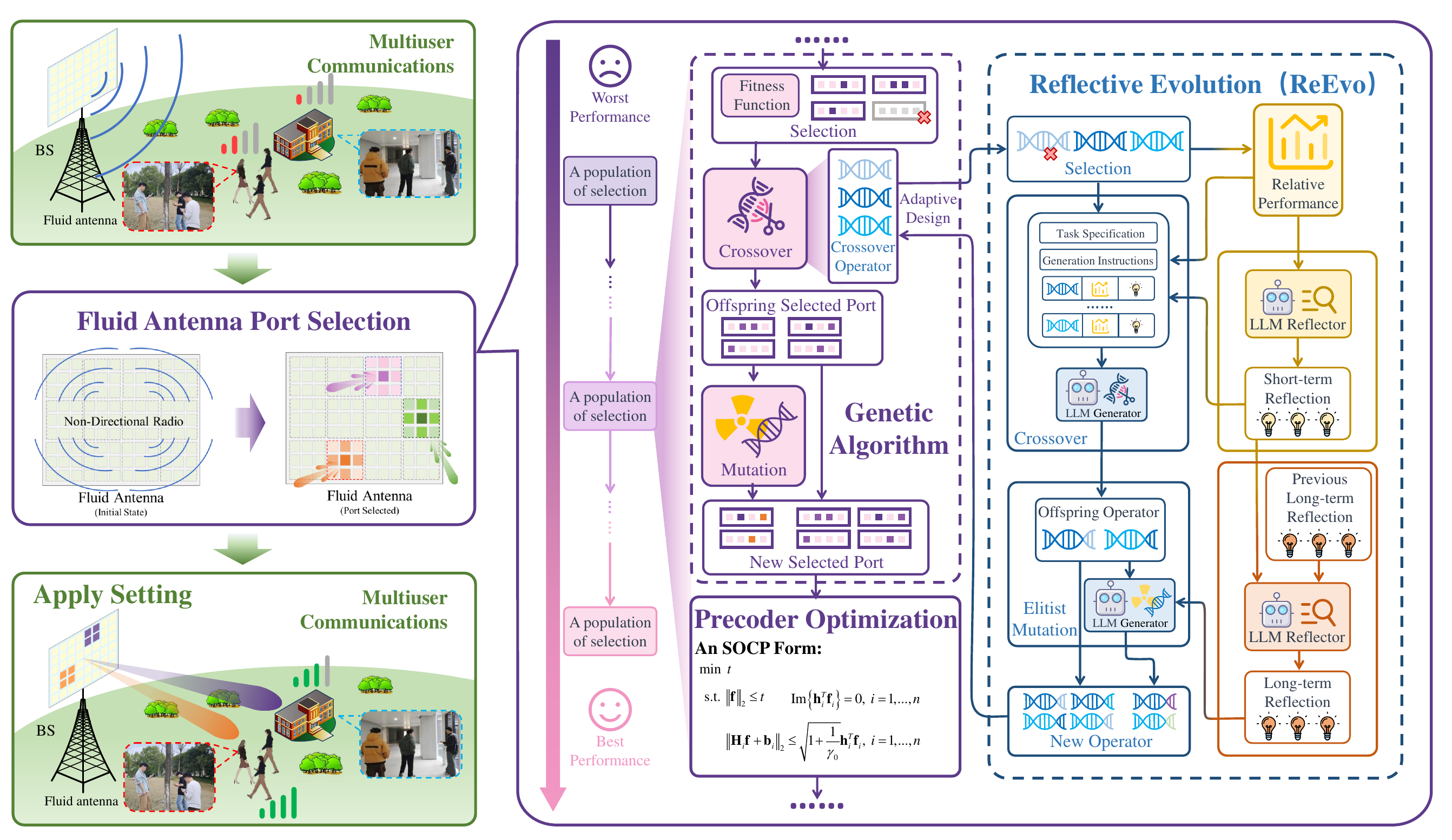}
	\caption{ReEvo-enhanced heuristic algorithm for \gls{FAS} design.}\label{CSIestimation}
	\vspace{-3mm}
\end{figure*}

\subsubsection{Flexible Precoder Design} %The design of \gls{FAS} precoders faces significant challenges due to the joint optimization of antenna positions and precoders across diverse scenarios. 
%These challenges are further compounded by the fact that the spatial correlation of \gls{FAS} ports varies with different scenarios, necessitating the redesign of both antenna positions and precoders for each case.
%Scene information can be acquired using vision sensors and emerging \gls{ISAC} technology. This sensed information can then be transformed into actionable command data through multimodal \gls{llm} and \gls{RAG}. 
%Here, \gls{RAG} extracts key system parameters from a knowledge base by leveraging the information generated by the multimodal \gls{llm}. Consequently, \gls{llm} can automate the formulation of design problems using this command information.
%However, unlike traditional array antenna precoder optimization, the joint design of fluid antennas constitutes an NP-hard combinatorial optimization problem. The inherent complexity and diversity of such problems often require domain specialists to develop heuristic methods for approximate solutions \cite{JointOptimizationMA}. 
%In this context, \gls{llm} offers new opportunities for automating heuristic design by exploring the knowledge base to identify the most effective heuristics.

Designing \gls{FAS} precoders is challenging due to the joint optimization of antenna positions and precoders across varying scenarios. Scene information, obtained via vision sensors or \gls{ISAC}, can be processed into actionable commands using multimodal \gls{llm} enhanced by \gls{RAG}. This approach automates heuristic design by leveraging knowledge bases to identify optimal strategies.

\textbf{\gls{llm} Enabled Automated Optimization Modeling:} Historically, \gls{FAS} design has depended on human expertise for problem modeling. 
%, due to the complexity of design objectives and the diversity of application scenarios. 
Recent breakthroughs in multimodal \gls{llm} technology, however, are reshaping this paradigm. These models can directly process multimodal sensor inputs (e.g., vision, radar) and generate corresponding mathematical representations, enabling end-to-end automated optimization. By systematically exploring the solution space, they accelerate the identification of high-performance designs across diverse operational conditions.
%While challenges such as hallucinations and slow knowledge updates currently limit the use of \gls{llm} in automated optimization modeling, these can be mitigated through \gls{RAG}. 
{\color{chao}
While hallucinations may limit the application of \gls{llm}s in \gls{FAS} design modeling, this issue can be alleviated through \gls{RAG} techniques. By augmenting \gls{llm}s with \gls{RAG}, they can integrate up-to-date domain knowledge from technical databases, enhancing accuracy and adaptability in \gls{FAS} model formulation. This approach bridges the gap between sensory data and optimized precoder design, enabling autonomous FAS development.}

\textbf{Automated Heuristic Design with \gls{llm}:} The design of \gls{FAS} requires simultaneous consideration of both antenna positioning and precoder optimization, a problem that is inherently NP-hard and thus challenging to solve optimally. Traditional methods often fall short in providing efficient solutions due to their reliance on predefined heuristic spaces shaped by human expertise. 
Recently, \gls{LHH}, an emerging automated heuristic design framework, has been introduced, utilizing \gls{llm} to generate heuristics~\cite{ye2024reevo}.
{\color{chao}The work of \cite{ye2024reevo} employs \gls{llm} to implement \gls{ReEvo}, a method that leverages the reasoning capabilities of \gls{llm} to mimic human expertise. This approach guides heuristic generation through iterative reflection on performance metrics. Specifically, \gls{ReEvo} features a dual-layer LLM agent architecture, equipping the LLM with both generative and reflective functionalities. By incorporating mechanisms such as selection, short-term reflection, crossover, long-term reflection, and elite mutation, \gls{ReEvo} effectively explores an open heuristic search space while continuously self-optimizing based on evolutionary feedback. This makes it particularly advantageous for optimizing heuristic algorithms like \gls{GA} in the context of \gls{FAS} joint design.
%\gls{GA} is widely recognized for its potential in solving combinatorial optimization problems, making it well-suited for handling the NP-hard joint design problem of \gls{FAS}. However, its effectiveness is often limited by the quality of its core operational components, such as selection, crossover, and mutation. Traditionally, these components require manual design and tuning for specific problem domains, which can lead to issues like premature diversity loss and convergence to local optima, hindering overall performance.
%To overcome these limitations, \gls{ReEvo}, as proposed by \cite{ye2024reevo}, can be applied to optimize \gls{GA}.
Applied to GA optimization, ReEvo treats genetic operation code snippets as evolutionary individuals and iteratively refines their design and implementation through multiple evolution cycles.} 
%This process not only enhances the performance of \gls{GA} but also addresses the constraints imposed by manual design, enabling more robust and adaptive solutions for \gls{FAS} optimization.

\begin{figure*}[!th]
	\centering
	\includegraphics[width=5.3in]{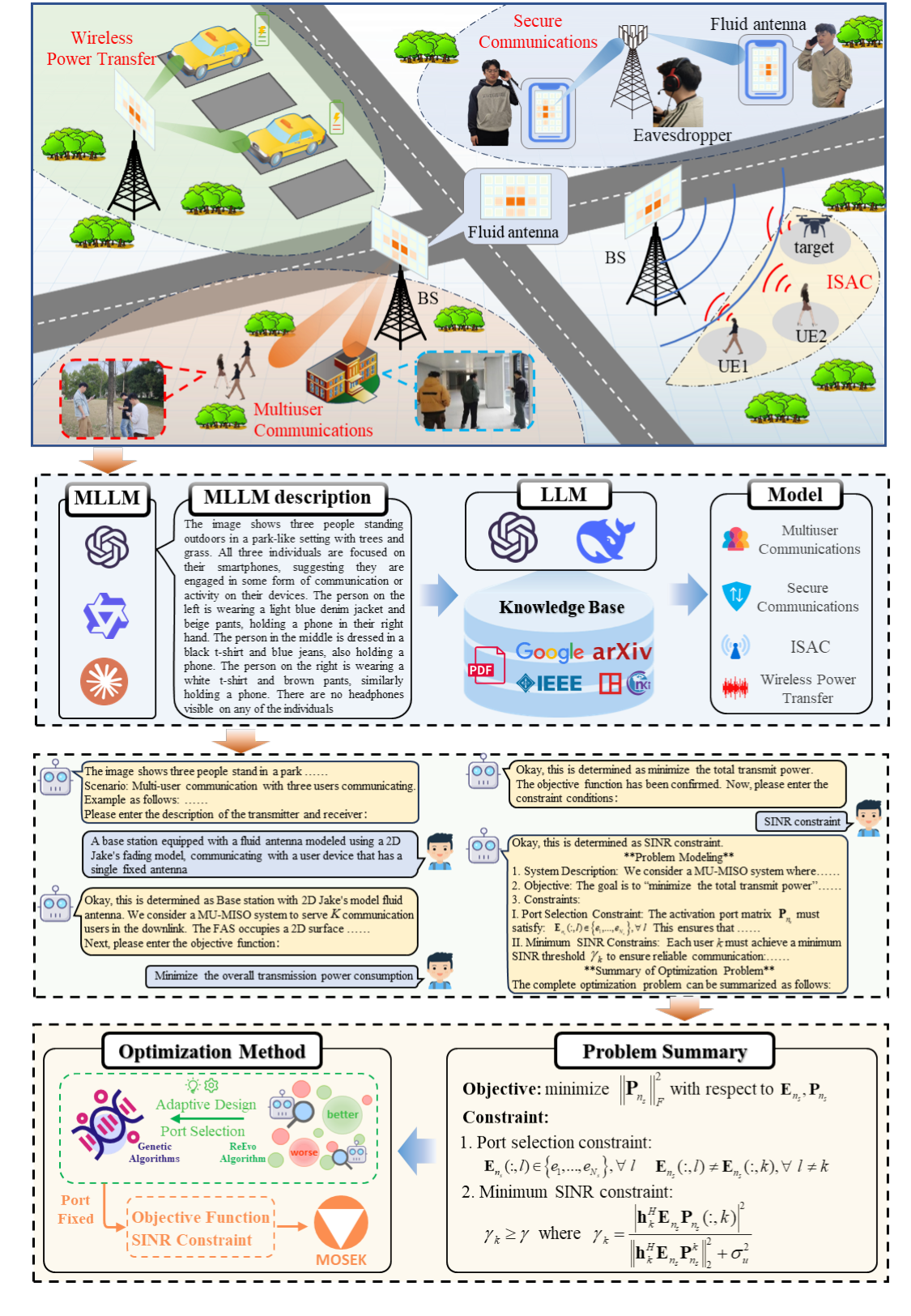}
	\caption{\gls{llm}-enabled intelligent \gls{FAS}.}\label{Fig:llm_FAS}
\end{figure*}

\subsubsection{Cooperative \gls{FAS}} Cooperative \gls{FAS} can significantly expand the design space by leveraging the spatial degrees of freedom from multiple nodes. However, this expansion introduces high complexity, rendering existing optimization approaches inadequate.
%To maximize cooperative gains, the antenna positions of multiple \gls{FAS} nodes must first be optimized to establish favorable cooperative channels. Subsequently, resource allocation strategies can be fine-tuned based on these optimized channels. As a result, cooperative \gls{FAS} systems exhibit significantly higher complexity compared to traditional cooperative communication systems.
%Additionally, each \gls{FAS} agent should possess self-configuration capabilities tailored to its role in the cooperative task. For instance, in collaborative multi-hop routing scenarios, FAS agents must dynamically optimize their configurations based on their local environment and the needs of target users.
%Furthermore, an efficient collaborative learning framework is essential to train multiple \gls{FAS} agents to work together effectively. Traditionally, wireless network optimization has relied heavily on domain-specific expert knowledge, which combines engineers' experience and logical reasoning.
\gls{llm}s, with their vast knowledge and superior inferential capabilities, provide a promising foundation for a knowledge-driven cooperative \gls{FAS} framework. By harnessing the collective intelligence of \gls{llm}-powered agents, autonomous cooperative \gls{FAS} can be achieved through cognitive synergy, enabling more efficient and adaptive systems.

\textbf{Autonomous Cooperative \gls{FAS}:} We propose an autonomous cooperative \gls{FAS} system driven by \gls{llm}-based agents, structured around four core components: (1) cooperative sensing, (2) task scheduling, (3) shared memory, and (4) self-examination. The sensing module leverages multimodal \gls{llm}s to generate rich, interpretable descriptions of the wireless environment, providing critical inputs for system optimization. 
The task scheduling module decomposes complex design problem into simpler sub-tasks and leverages the \gls{CoT} technique to guide \gls{llm} decision-making. 
%Specifically, \gls{CoT} breaks down intricate problems into smaller, logical steps, enabling structured reasoning and incremental problem-solving for accurate solutions. 
The shared memory module utilizes a common vector database to store system design experiences and expert knowledge from public data. {\color{chao}Each \gls{FAS} agent can then employ \gls{RAG} to retrieve relevant insights from this shared memory to support its decisions.} Finally, the reflection module empowers \gls{FAS} agents to emulate human-like cognitive learning, iteratively improving decision-making by analyzing historical performance data, extracting insights from both successes and failures.

 \begin{figure}[!t]
	\centering{\includegraphics[width=0.97\columnwidth]{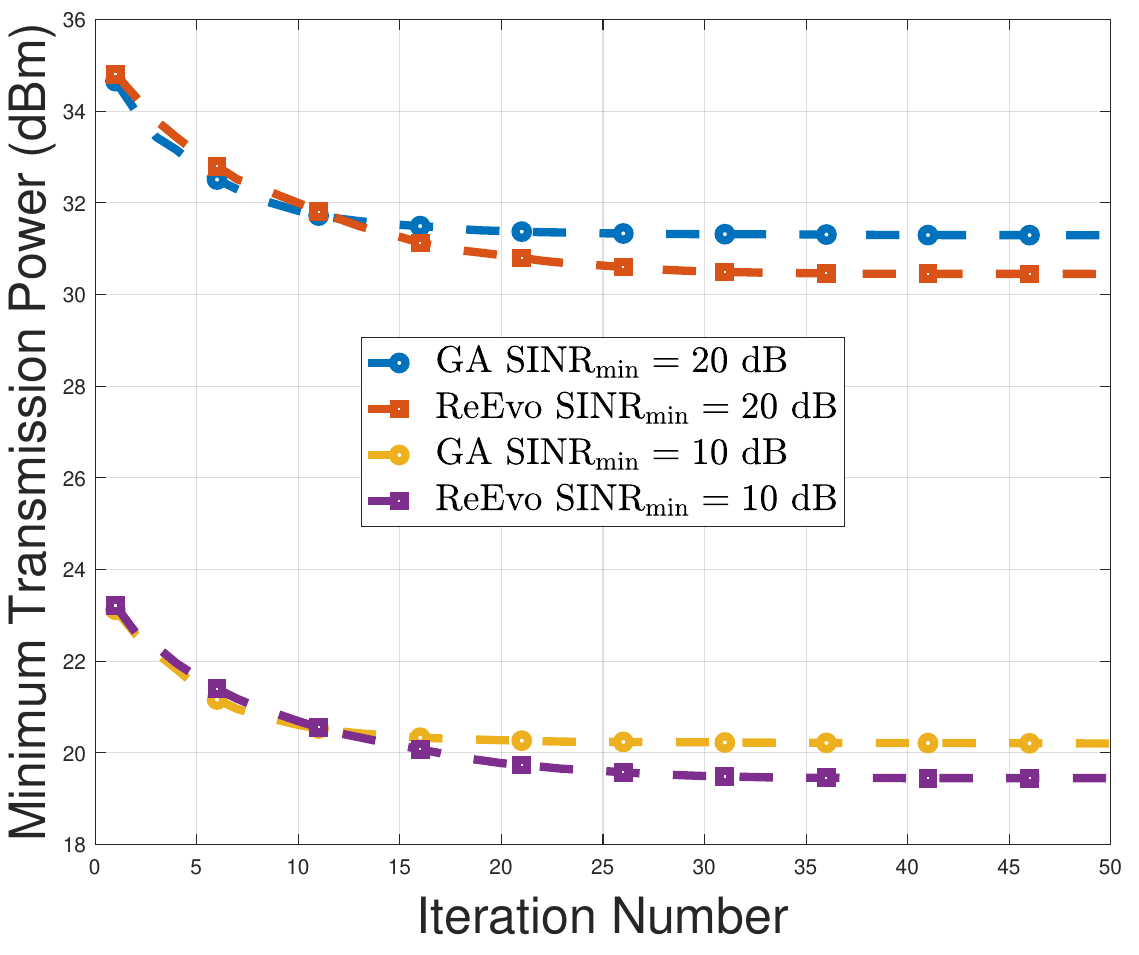}}
	\caption{Average Minimum Downlink Transmission Power versus Iteration Number in a 10 single-antenna User FAS System (0 dBm user noise power, small-cell \gls{BS} FAS: 10 cm $\times$ 10 cm, 10 $\times$ 10 positions, 10 activated ports)}\label{TransmittedVSIteration}
		\vspace{-3mm}
\end{figure}

\textbf{\gls{llm}-Based Multi-Agent Reinforcement Learning:}  \gls{MARL} has been extensively employed to develop autonomous agents capable of learning cooperative behaviors in complex environments. As a result, \gls{MARL} can be effectively applied to train cooperative strategies for multiple \gls{FAS} agents. However, a major challenge in such systems is the ``curse of dimensionality," which arises due to the large number of \gls{FAS} agents. Specifically, the joint design of antenna positioning and precoding leads to an exponential expansion of the state and action spaces, causing issues such as algorithmic instability, difficulties in convergence, and a higher likelihood of becoming stuck in local optima. To address these challenges, it is crucial to develop an efficient MARL training algorithm tailored for cooperative FAS systems. \cite{MARL_Training} proposed \gls{eSpark}, a framework that leverages \gls{llm}'s prior knowledge and encoding capabilities to improve exploration efficiency. \gls{eSpark} operates through three core mechanisms: (1) generating diverse exploration functions from sensing data using \gls{llm} as a code generator, (2) searching for the optimal \gls{MARL} policy by evaluating and sampling high-performing functions, and (3) refining exploration functions through policy feedback, where environmental rewards guide iterative improvements to the \gls{llm}'s generation process.
\begin{Remark}
{\color{chao}While LLMs offer transformative potential for automated heuristic design in \gls{FAS}, their practical deployment depends on {model compression}, {hardware-aware optimization}, and {latency reduction}. Although foundational \gls{llm} is typically large, {distilled or sparse models} (e.g., \href{https://tinyllm.org}{TinyLLM}\footnote{\url{https://tinyllm.org}}) can replicate their behavior while retaining most of the performance at {significantly lower latency}. To address hardware constraints, \gls{llm} can be trained in {cloud environments} while deploying inference on {edge-optimized accelerators}. For flexible precoder design, \gls{fpga}/\gls{asic} implementations can execute fixed heuristics after \gls{llm}-guided optimization.
Further reducing \gls{llm} inference latency can be achieved through a {hierarchical agent architecture}:
\begin{itemize}
	\item A fast \textbf{``Actor"} (a lightweight neural network for real-time decisions)
	\item A slow \textbf{``Reflector"} (a larger \gls{llm} refining strategies offline and updating the Actor periodically)
\end{itemize}
Additionally, {specialized hardware}---such as \gls{gpu}, \gls{tpu}, or \gls{fpga} optimized for DL workloads---can further accelerate inference speeds.}
\end{Remark}

\begin{figure}[!t]
	\centering{	\includegraphics[width=1.0\columnwidth]{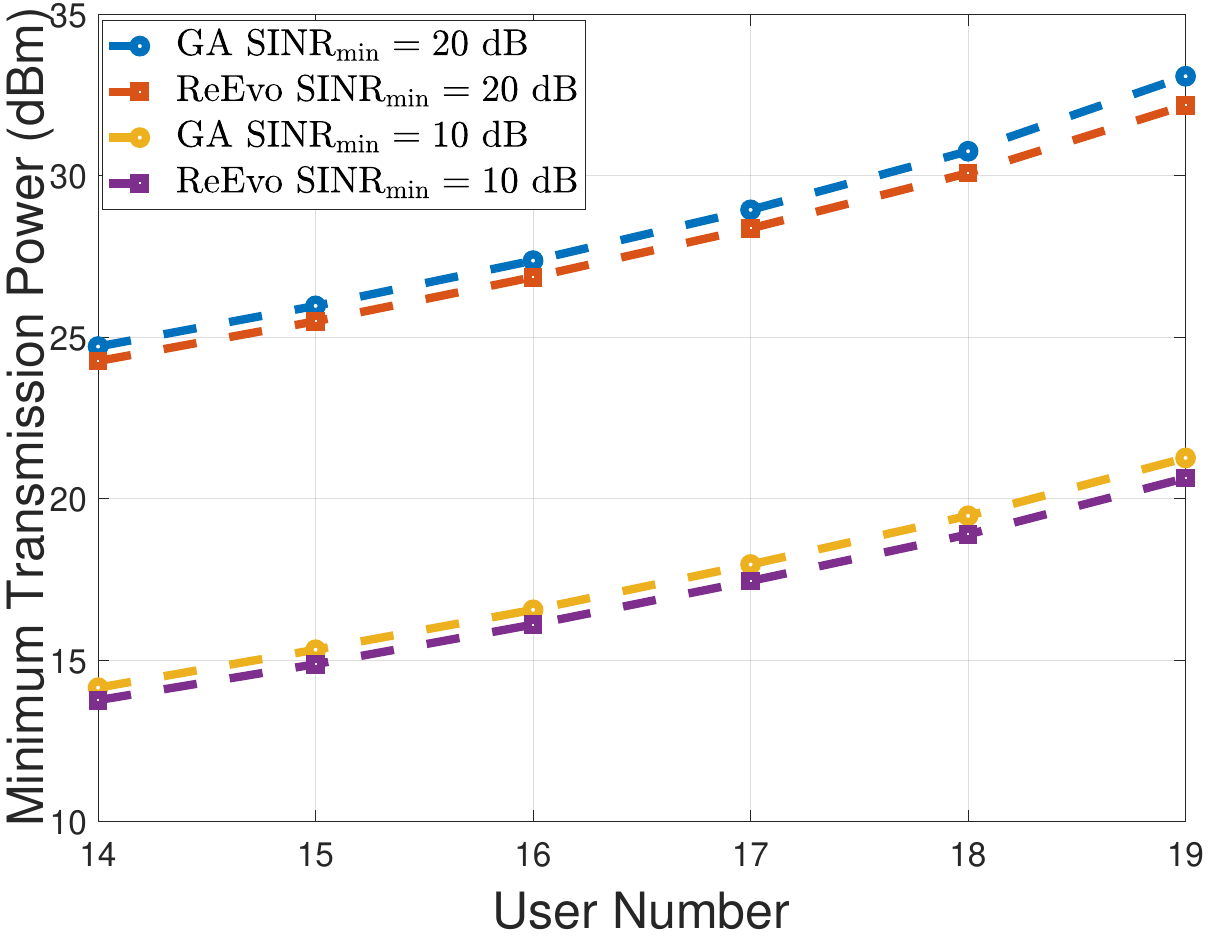}}
	\caption{Average Minimum Downlink Transmission Power versus $K$ in a FAS System (0 dBm user noise power, small-cell \gls{BS} equipped with FAS: 20 cm $\times$ 20 cm, 10 $\times$ 10 positions, 20 activated ports).}\label{TransmittedVSNumberofUser}
		\vspace{-3mm}
\end{figure}

\begin{Remark}
	{\color{chao}
	While RAG-enhanced LLMs improve knowledge retrieval and automate modeling for FAS design, their reliance on probabilistic reasoning introduces critical limitations, most notably, hallucinations. In FAS applications, where precise modeling of physical phenomena (e.g., objective functions, beamforming constraints) is essential, LLMs may produce plausible but factually inconsistent output.
	To ensure the validity of an LLM-generated optimization problem, implement a multi-step verification process combining automated checks and LLM-assisted analysis. First, verify the problem's well-posedness by confirming its objective function, variables, and constraints are properly defined and logically consistent. Next, leverage optimization solvers (e.g., CVXPY, Gurobi) to automatically assess feasibility, boundedness, and constraint satisfaction. The LLM can then refine the formulation by cross-examining semantic alignment with the original intent, identifying constraint conflicts, and suggesting tractable reformulations. To mitigate potential LLM hallucinations, incorporate RAG to infuse domain-specific knowledge for critical applications. This hybrid approach, integrating solver-based validation, LLM verification, and expert oversight, strikes an optimal balance between robustness and scalability in optimization problem validation.}
\end{Remark}

\section{Case Study: \gls{llm}-Enabled Intelligent \gls{FAS}}\label{sec:isac}
In this section, we leverage QwenLM\footnote{\url{https://github.com/QwenLM/Qwen}}, a widely used \gls{llm}, to propose an intelligent design framework for \gls{FAS}. 
%Specifically, designing an intelligent \gls{FAS} consists of five aspects including environment modeling, transmission protocol, optimization goal, channel model, and optimization algorithm design.   
We first utilize Qwen-VL-Max\footnote{\url{https://github.com/QwenLM/Qwen-VL}}, a well-known \gls{mllm}, to build the problem model from the visual information of the scenario. Next, we introduce a \gls{ReEvo}-enhanced \gls{GA} to address the combinatorial optimization challenges inherent in \gls{FAS}.

\subsection{Problem Automated Modeling Based on Vision Information}
An intelligent FAS should generate tailored designs for diverse applications by first extracting key parameters from scene information to model the problem and autonomously determine the appropriate solution. This article explores four key FAS applications: multiuser communications, secure communications, ISAC, and energy transfer. 
%Each application involves distinct types of users. For instance, secure communications typically involve eavesdroppers equipped with headsets, while \gls{ISAC} includes scenarios with cars or unmanned aerial vehicles. 
Fig. \ref{Fig:llm_FAS} focuses on multiuser communications\footnote{Detailed tutorials on the other case studies are available at \url{https://drwangchao.github.io/LLLMDOFAS/}}.

To enable automated optimization modeling, three agents are employed: the extraction agent, the terminology agent, and the problem description agent. The extraction agent uses Qwen-VL-Max to extract scene information from visual perception. As shown in Fig. \ref{Fig:llm_FAS}, the extraction agent generates a textual description of the scene, such as, ``The image shows three people standing outdoors in a park-like setting with trees and grass\dots'' The terminology agent then translates this description into a terminological format by referencing a knowledge base containing problem models for the four applications, i.e., ``There are three users in a multiuser communications scenario.'' {\color{chao}The problem description agent then uses the \gls{RAG} technique to perform a semantic similarity search, identifying the appropriate problem model for multiuser communications. }
%Specifically, the problem description comprises three components: the description of the transmitter and receiver, the objective function, and the system constraints. Based on this model, users can input the necessary information to construct the corresponding mathematical problem.

Next, leveraging QwenLM, the system uses structured guidance to build the mathematical problem model. For each component in the problem description, the structured guidance matches user queries with specific \gls{FAS} modeling knowledge. In Fig. \ref{Fig:llm_FAS}, the user inputs a query for the objective function, ``minimize the overall transmission power consumption," and the system applies RAG techniques to extract relevant objective functions from IEEE Explore. The problem description includes transmitter and receiver details (e.g., ``A base station equipped with a fluid antenna modeled using a 2D Jake's fading model...."), the system automatically generates optimization variables (e.g., port selection and broadcasting precoder) along with the necessary constraints.
Using this structured guidance, a detailed optimization problem is constructed, including the objective function, constraints, and optimization variables. This problem can then be solved using the ReEvo-enhanced heuristic algorithm.

In conclusion, \gls{llm} can automate problem modeling and solving, enabling an intelligent implementation framework for \gls{FAS}. This approach streamlines the design and optimization process, making \gls{FAS} more adaptable and efficient for diverse applications. 
\subsection{\gls{ReEvo}-Enhanced Heuristic Algorithm for \gls{FAS} Design}

Through automated problem modeling, we formulate a joint optimization of port selection and precoder design to minimize \gls{bs}'s transmit power while satisfying user SINR constraints (SINR$_{\min}$). Due to the nonconvex nature of this problem, we employ an alternating optimization approach that decomposes it into two tractable subproblems:
Port selection optimization (combinatorial problem) and Precoder design (convex problem).
{\color{chao}While conventional GA effectively handles combinatorial optimization through population diversity and genetic operations, its performance remains limited by the expert-dependent design of crossover operators. As shown in Fig. \ref{CSIestimation}, our proposed ReEvo framework overcomes this limitation through an intelligent multi-level optimization process:
1) Initialization: A generation LLM creates initial crossover operator code snippets;
2) Selection and Analysis: High-performing parents are selected and analyzed by a reflection \gls{llm};
3) Offspring Generation: The generation LLM produces improved offspring by combining parental traits with reflection insights;
4) Refinement: Long-term reflection and elite mutation further enhance operator performance.
Unlike traditional metaheuristics (e.g., genetic algorithms), \gls{ReEvo} achieves superior scalability compared to traditional methods in complex search spaces with minimal manual parameter tuning, thanks to its self-improving heuristic adaptation. By dynamically refining its search process, ReEvo enables faster convergence and higher-quality solutions for the target problem\footnote{We have given detailed \href{https://drwangchao.github.io/LLLMDOFAS/static/pdfs/ReEvo.pdf}{dialogue}, \href{https://drwangchao.github.io/LLLMDOFAS/static/pdfs/LLM\%20Prompt.pdf}{prompt}, and \href{https://drwangchao.github.io/LLLMDOFAS/static/pdfs/port\_selection-ga.log}{running log} for ReEvo-enhanced heuristic algorithm. }
}.
After port selection, the precoder design is transformed into a second-order cone programming (SOCP) problem, efficiently solved using convex optimization solvers.

{\color{chao}To demonstrate the potential of \gls{llm} in heuristic design, we compare the performance of ReEvo-enhanced GA (labeled ReEvo) with that of traditional GA (labeled GA) in Figs.~\ref{TransmittedVSIteration} and~\ref{TransmittedVSNumberofUser}.
	The \gls{FAS} channel follows a spatially correlated Rayleigh fading model, with inter-port correlation modeled using a 2D Jake's approach based on the zeroth-order spherical Bessel function. The carrier frequency is 3.4 GHz, path loss is neglected, and the received noise is normalized to 0 dBm. The GA uses a population size of 200, an elite rate of 20\%, and runs for 50 generations.} As expected, \gls{ReEvo} achieves significantly lower transmitted power, particularly for large user counts ($K$).
Fig.~\ref{TransmittedVSIteration} illustrates the average convergence rates of ReEvo and GA. Notably, ReEvo achieves a {17.6\% improvement} over GA at $\text{SINR}_{\min} = 20~\text{dB}$ and a {16\% gain} at $\text{SINR}_{\min} = 10~\text{dB}$. Similarly, Fig.~\ref{TransmittedVSNumberofUser} compares the minimum transmission power as $K$ increases, where ReEvo outperforms GA by {12.8\%} ($20~\text{dB}$) and {10\%} ($10~\text{dB}$).
These results validate the {superior efficiency of \gls{llm}} in optimizing heuristic algorithms for \gls{FAS} design.

\section{Conclusions and Future Research}\label{sec:con}

This work explored the integration of \gls{llm} into the design and optimization of \gls{FAS}, demonstrating its transformative potential to accelerate innovation in wireless communications. Unlike conventional DL methods, \gls{llm}s offer unique advantages for efficient \gls{FAS} design, particularly through their ability to capture long-sequence dependencies, a capability we leverage for \gls{CSI} prediction. Furthermore, \gls{llm}s enable automated problem modeling for diverse \gls{FAS} applications and facilitate the development of heuristic algorithms to enhance system performance.
Beyond optimization, we proposed an \gls{llm}-empowered autonomous cooperative \gls{FAS} framework capable of cooperative sensing, task scheduling, and self-directed learning. To illustrate the practical impact of this approach, we applied \gls{llm}-driven solutions to multiuser \gls{FAS} communications, showcasing their adaptability and performance gains.
This research underscores \gls{llm}s as a paradigm-shifting tool for \gls{FAS} design, bridging theoretical modeling with real-world deployment challenges. Future work will focus on hardware validation and domain-specific \gls{llm} fine-tuning to further unlock their potential. Based on the insights from this article, {\color{chao} we propose the following future research directions:

\textbf{NOMA Meets \gls{FAS}}: \gls{NOMA} is widely regarded as a potential technology for 6G networks, primarily due to its superior spectral efficiency and ability to support massive connectivity. Unlike conventional orthogonal multiple access (OMA) schemes, NOMA allows multiple users to share identical time-frequency resources by superimposing their signals through power-domain or code-domain multiplexing. However, integrating NOMA with FAS introduces unique challenges, as traditional NOMA protocols assume static channel conditions. In FAS, where channel characteristics are highly dynamic due to reconfigurable antenna ports, key NOMA mechanisms, such as power allocation and user pairing, must be continuously adapted. Specifically, power allocation must account for rapidly fluctuating channel gains, while user clustering strategies need to dynamically adjust to varying spatial correlations induced by port switching. Consequently, the port selection process in FAS must be redesigned to align with NOMA's adaptive requirements, ensuring optimal performance in dynamic wireless environments.

\textbf{RIS-Aided \gls{FAS}}: \gls{IRS} is poised to revolutionize 6G wireless networks by dynamically optimizing the propagation environment. The integration of \gls{IRS} and FAS offers transformative potential for 6G, enabling unprecedented control over wireless channels. However, this synergy also introduces critical research challenges. Specifically, both \gls{IRS} and FAS rely on real-time channel adaptation, yet their combined dynamics, such as FAS's port switching and IRS's beamforming adjustments complicate \gls{CSI} estimation. The first key challenge is acquiring accurate \gls{CSI} for joint optimization. Additionally, co-designing \gls{IRS} phase shifts and FAS port selection under rapidly varying channels demands efficient, low-complexity algorithms.}

\ifCLASSOPTIONcaptionsoff
  \newpage
\fi


\bibliographystyle{IEEEtran}
%\bibliography{reference}

\begin{thebibliography}{10}
\providecommand{\url}[1]{#1}
\csname url@samestyle\endcsname
\providecommand{\newblock}{\relax}
\providecommand{\bibinfo}[2]{#2}
\providecommand{\BIBentrySTDinterwordspacing}{\spaceskip=0pt\relax}
\providecommand{\BIBentryALTinterwordstretchfactor}{4}
\providecommand{\BIBentryALTinterwordspacing}{\spaceskip=\fontdimen2\font plus
\BIBentryALTinterwordstretchfactor\fontdimen3\font minus
  \fontdimen4\font\relax}
\providecommand{\BIBforeignlanguage}[2]{{%
\expandafter\ifx\csname l@#1\endcsname\relax
\typeout{** WARNING: IEEEtran.bst: No hyphenation pattern has been}%
\typeout{** loaded for the language `#1'. Using the pattern for}%
\typeout{** the default language instead.}%
\else
\language=\csname l@#1\endcsname
\fi
#2}}
\providecommand{\BIBdecl}{\relax}
\BIBdecl

\bibitem{Tutorial_FAS}
W.~K. New {\em et al.}, ``A tutorial on fluid antenna system for {6G} networks: Encompassing communication theory, optimization methods and hardware designs,'' {\em IEEE Commun. Surv. \& Tut.}, \url{doi: 10.1109/COMST.2024.3498855}, 2024.

\bibitem{VirtualFAS}
K.-K. Wong {\em et al.},  ``Virtual {FAS} by learning-based imaginary antennas,'' \emph{IEEE Wireless Commun. Lett.}, vol.~13, no.~6, pp. 1581--1585, Jun. 2024.

\bibitem{ISAC_FAS}
C.~Wang {\em et al.}, ``Fluid antenna system liberating multiuser {MIMO} for {ISAC} via deep reinforcement learning,'' \emph{IEEE Trans. Wireless Commun.}, vol.~23, no.~9, pp. 10\,879--10\,894, Sept. 2024.

\bibitem{AI_FAS}
C.~Wang {\em et al.}, ``{AI}-empowered fluid antenna systems: {Opportunities}, challenges, and future directions,'' \emph{IEEE Wireless Commun.}, vol.~31, no.~5, pp. 34--41, Oct. 2024.

\bibitem{Niyato1}
J.~Wang {\em et al.},  ``Generative {AI} enabled robust data augmentation for wireless sensing in {ISAC} networks,'' \emph{arXiv preprint}, \url{arXiv:2502.12622}, Feb. 2025.

\bibitem{Niyato2}
R.~Zhang {\em et al.}, ``Generative {AI} agents with large language model for satellite networks via a mixture of experts transmission,'' \emph{IEEE J. Sel. Areas Commun.}, vol.~42, no.~12, pp. 3581--3596, Dec. 2024.

\bibitem{FAS-LLM1}
H.~Yang, Zhao, S.~Lambotharan, and M.~Derakhshani, ``{FAS-LLM}: Large language model-based channel prediction for {OTFS}-enabled satellite-{FAS} links,'' \emph{arXiv preprint}, \url{arXiv:2505.09751}, May 2025.

\bibitem{Port_LLM}
Y.~Zhang, H.~Yin, W.~Li, E.~Bj\"{o}rnson, and M.~Debbah, ``{Port-LLM}: {A} port prediction method for fluid antenna based on large language models,'' \emph{arXiv preprint}, \url{arXiv:2502.09857}, Feb. 2025.

\bibitem{CKM_ChannelPrediction}
X.~Wang {\em et al.}, ``Channel knowledge map aided channel prediction with measurements-based evaluation,'' \emph{IEEE Trans. Commun.}, vol.~73, no.~5, pp. 3622--3636, May 2025.

\bibitem{MARL_Training}
Z.~Liu {\em et al.}, ``Knowing what not to do: {Leverage} language model insights for action space pruning in multi-agent reinforcement learning,'' \emph{arXiv preprint}, \url{arXiv:2405.16854}, May 2024.

\bibitem{mllm}
S.~K. {Yin} {\em et al.}, ``A survey on multimodal large language models,'' \emph{arXiv preprint}, \url{arXiv:2306.13549}, Nov. 2024.

\bibitem{RAGLLM}
J.~Chen, H.~Lin, X.~Han, and L.~Sun, ``Benchmarking large language models in retrieval-augmented generation,'' \emph{Proc. of AAAI Conf. Artificial Intelligence}, vol.~38, no.~16, pp. 17\,754--17\,762, Dec. 2024.

\bibitem{zhang2023using}
M.~Zhang, N.~Desai, J.~Bae, J.~Lorraine, and J.~Ba, ``Using large language models for hyperparameter optimization,'' in \emph{Proc. NeurIPS Foundation Models for Decision Making Workshop}, Dec. 2023.

\bibitem{LLM_channelestimation}
B.~Liu, X.~Liu, S.~Gao, X.~Cheng, and L.~Yang, ``Llm4cp: Adapting large language models for channel prediction,'' \emph{J. Commun. Inf. Netw.}, vol.~9, no.~2, pp. 113--125, Jun. 2024.

\bibitem{ye2024reevo}
H.~Ye {\em et al.}, ``Reevo: Large language models as hyper-heuristics with reflective evolution,'' in \emph{Proc. Adv. Neural Inf. Process. Syst.}, Dec. 2024, \url{https://github.com/ai4co/reevo}.
\end{thebibliography}

\end{document}